# Design of a mechanically actuated RF grounding system for the ITER ICRH antenna

David Hancock[a], Mark Shannon[a], Bertrand Beaumont[b], Pierre Dumortier[c], Frederic Durodie[c], Volodymyr Kyrytsya[c], Fabrice Louche[c], Robert McKinley[a], Keith Nicholls[a], and the CYCLE Team[a,c,d,e,f]

[a]EURATOM/CCFE Fusion Association, Culham Science Centre, Abingdon, OX14 5ER, UK
[b]ITER Organisation, Route de Vinon sur Verdon, 13115 Saint Paul Lez Durance, France
[c]LPP/ERM-KMS, Association EURATOM-Belgian State, Brussels, Belgium
[d]Association EURATOM-CEA, DSM/IRFM, Cadarache, France
[e]Max-Planck Institut fur Plasmaphysik, EURATOM-Assoziation, Garching, Germany
[f]Associazione EURATOM-ENEA, Politecnico di Torino, Torino, Italy

In the ITER equatorial ports containing ICRH antennas, parasitic electrical resonances can be excited in the nominal 20 mm clearance gap between the port walls and the plug contained within it. RF calculations have established that these resonances can be effectively mitigated by a series of suitably located electrically conducting contacts between the port and plug. These contacts must allow installation and removal of the antenna but must also make reliable electrical contact during antenna operation. In addition, the contacts must be compliant enough to survive deflection of the port during seismic and disruption events without transmitting large forces to the vacuum vessel. The distance to be spanned is subject to significant uncertainty, due to the large manufacturing tolerances of the surrounding components, and this also must be considered during the design process. This paper outlines progress made in the design of the grounding system, as well as detailing a number of concepts which have been investigated and abandoned, leading up to the current reference design. The current reference design is a simple and robust mechanical solution consisting of sprung Copper-plated Inconel flaps which use part of the range of the shimming system included in the antenna design as the actuation mechanism. This paper also details practical testing of a number of aspects of the design, including building and testing a prototype to validate mechanical and thermal analyses.

## 1. Introduction

In the ITER equatorial ports, a nominal 20 mm clearance gap exists between the port walls and the plug contained within it. RF calculations have established that in the ports containing an ICRH antenna, parasitic electrical resonances in this gap at ICRH frequencies can be excited which have an adverse effect on the performance of the antenna and induce large voltages on surrounding components. Further calculation has shown that these resonances can be effectively mitigated by a series of electrically conducting contacts between the port and plug, recessed 1m from the first wall [1]. In order to be able to install and remove the antenna from the port, these contacts must not interfere with the surrounding components while still being able to make reliable electrical contact once the antenna is in operation. In addition, the contacts must be compliant enough to survive deflection of the port during seismic and disruption events without transmitting large forces to the vacuum vessel. The 20 mm distance to be spanned is subject to significant uncertainty, due to the large manufacturing tolerances of the surrounding components, and this also must be considered during the design process.

## 2. Preliminary concept development

Prior to settling on the current design outlined below, a wide range of concepts were explored. While precise RF geometry requirements were not available, general considerations, such as continuous contact surfaces, drove the design from an early stage. The first RF analyses [1] modelled a series of cylindrical "buttons" as shown in figure 1, and hence the early designs followed this pattern.

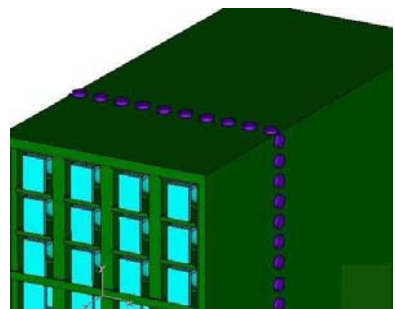

Fig. 1: RF model of grounding contacts from [1]

It was anticipated that, in the interest of limiting disruption currents, the contacts could be made capacitive rather than galvanic using ceramic coating on the top surface rather than extending the conducting material throughout the contact. Modelling showed, however, that the amount of capacitance needed and the tolerances required rendered capacitive contacts impractical [2]. Permission was then obtained from the ITER Organisation to explore the effect of galvanic contacts. Subsequent disruption analyses have therefore

*Author's email: david.hancock@ccfe.ac.uk*

included fully conductive elements between the port extension and plug body.

It became obvious that, with the number of penetrations already present in the rear bulkhead and the increasingly stringent space requirements, employing a stand-alone, passive system would be of no small benefit to the antenna design as a whole, and so solutions were sought which employed systems or environmental conditions already present in the antenna design.

For the early concepts the current carrying surface was taken to be a cylinder of dimensions equal to those used for the modelling done in [1]. While sliding cans with sprung contact strips were explored, the RF surface used for the majority of these designs was a Copper plated bellows of appropriate diameter with an axial stroke equal to the gap to be bridged. Later concepts abandoned this constraint once further modelling showed that the effectiveness of the grounding was sufficiently retained with less "idealised" contacts [3].

## 2.1 Hydraulic and pneumatic concepts

With the ubiquity of cooling water throughout the antenna, including potentially within the buttons themselves, and the need to actively cool the contact, the possibility of employing coolant pressure to extend the bellows was explored, but abandoned due to both the potential vulnerability of the flexible surface and the limited motion available once a sufficiently robust double bellows was designed. The complexity of the additional water connections needed with the added difficulty of providing monitored interspaces was also a significant factor in this decision. Gas-pressurised options provided less risk to the machine, but suffered similar difficulties and were also ruled out.

## 2.2 Thermally actuated concepts

The antenna is to be installed at ambient temperature and raised to the temperature of the cooling water (70-126°C) during operation. This temperature difference provided the possibility of using a thermally actuated mechanism. Two distinct solutions presented themselves: bimetallic springs and shape memory alloys.

### 2.2.1 Bimetallic springs

A detailed parametric study of material pairings and bimetallic spring geometries was undertaken. In order to provide the range of motion within the constraint of the cylindrical envelope, spiral arrangements were used, some of which are shown in figure 2.

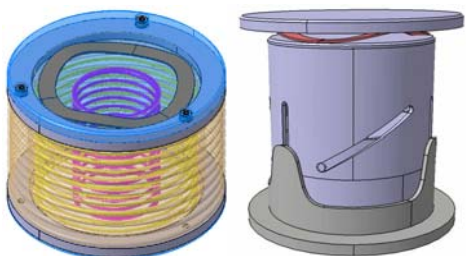

Fig. 2: Bimetallic concepts

The leading candidate spring geometry was accidentally discovered, Archimedes-like, as part of the thermostatic valve in the lead author's shower. This doubly-coiled spiral, as shown in figure 3, provides completely linear motion of the required range within the geometric envelope available.

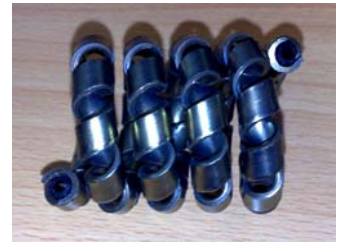

Fig. 3: Double spiral bimetallic spring

Unfortunately, the material combination of Invar and ferritic stainless steel used in this example is not suitable for use in ITER and other material combinations explored did not provide the required combination of actuation force and range. Further parametric FEA studies sought to improve the geometry and material combination, but these all fell short by approximately a factor of 10 in either stiffness or deflection. It is worth noting that these studies did not include the effects of remaining at elevated temperature during the duration of baking cycles or the effect of cyclic loading, and hence ignored potential creep and fatigue issues.

### 2.2.2 Shape Memory Alloys

A range of alloys known as shape memory alloys (SMAs), including the Nickel-Titanium alloy known as Nitinol, exhibit large strain recovery when raised above a critical temperature and provide much higher force/displacement ratios than bimetallic springs. Research showed, however, that time dependent effects caused by stress at elevated temperature reduced the shape memory effects too substantially for these to be used in the concepts proposed [4].

## 2.3 Static concepts

A universal installation method for the equatorial port plugs has yet to be specified and the possibility of employing custom-machined static blocks attached to both the port extension and plug wall which could meet with small degrees of compliance was examined. This method could be used if the plug were to be cantilevered horizontally into place without use of wheels or skids. The tolerances required as well as the risk of fouling surrounding components leaves this as an impractical solution with the installation methods proposed at this time.

## 2.4 Wedge operated concepts

Sprung contacts retracted using wedges removed post-installation were also explored. One example is shown in Figure 4. These required access to the front face of the antenna from the vessel or from the rear through the mounting flange, neither of which are now possible following subsequent geometry changes which were driven primarily by neutronic considerations. The number and complexity of sliding contacts required also left these concepts impractical.

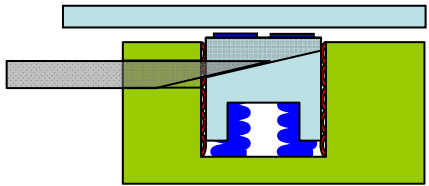

Fig. 4: Removable wedge operated concept

## 2.5 Use of the antenna shimming system

Hydraulic rams are used to move the internal components of the antenna radially within the port plug to optimise the distance between the radiating straps and the plasma [5]. Extending the range of this shimming allows part of this motion to be used to actuate the grounding. Using this pre-existing functionality avoids adding further penetrations to the bulkhead, and does not increase control requirements. Preliminary concepts maintained the bellows-based contact of their predecessors, but were quickly replaced by alternatives using a range of contact shapes. Wedge concepts similar to those previously developed and lever concepts employing flexible joints, levers and vacuum bearings were among those considered. Two such concepts are shown in Figure 5.

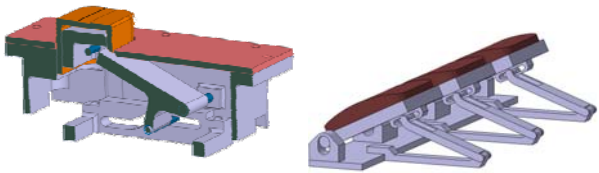

Fig 5: Lever and pedal shimming-operated concepts

These all, however, required sliding contacts and/or complex mechanisms which were deemed too vulnerable for this application.

## 3. Description of current reference design

### 3.1 Overview

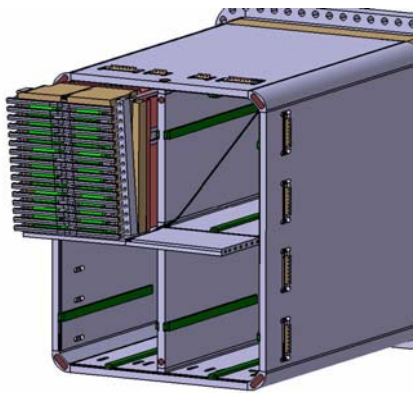

Fig 6: Contact layout

The current design of the grounding system consists of a series of copper-plated Inconel leaf springs attached to the port plug body and a corresponding series of staple-shaped crossbars, the legs of which pass through the port plug and are attached to the underlying structure. Figure 6 shows the location of these contacts, while figure 7 shows a close-up of a single contact mechanism.

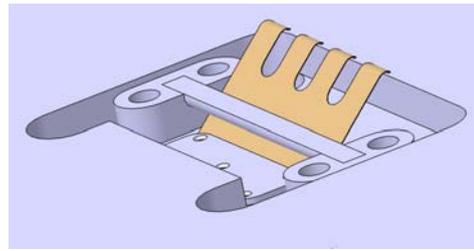

Fig. 7: Detail of grounding contact

### 3.2 Contact

The spring contacts are heat treated to achieve the required material properties before being electroplated with a minimum thickness of 100 μm of Copper to enhance electrical and thermal conductivity. Two sizes of spring are used, width constrained by the underlying structure, each with a finger-design to ensure multiple points of contact with the vessel. In order to allow remote handling replacement of the contact and to avoid high temperature bonding techniques which would damage the plating and could upset the spring temper of the Inconel, the contacts are bolted to the port plug. In order to improve the thermal and electrical conductivity of this bond, a thin layer of graphite gasket material, Papyex, is used.

### 3.3. Actuation method

Each contact is able to be depressed via a crossbar, the legs of which pass through the plug wall and are bolted to the underlying internal antenna components. The relative motion of these components and the port plug wall causes the crossbar to depress the contacts at the rearmost shimming position. The first 20mm of a total shimming range of 50mm are used for grounding actuation, as shown in Figure 8.

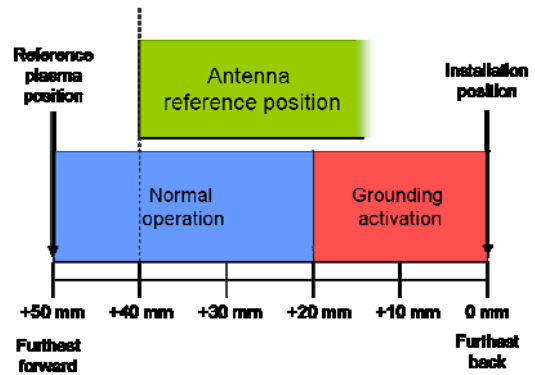

Fig 8: Shimming range

## 4. Analysis, prototype manufacture, and testing

Mechanical analyses using ANSYS optimised the spring dimensions and plating thickness to ensure contact force is maintained, while allowing flexibility without plastic deformation. Thermal analyses in parallel established the thickness of plating required to increase the thermal conductivity of the contact so that the heat from RF and neutronic heating could be effectively removed without additional water cooling. Figure 9 shows the results of a thermal analysis which included neutronic and RF heating and showed a maximum expected temperature of 250°C. RF heating was calculated based on pessimistic values and uniformly applied to contact surfaces. Neutronic heating assumed plasma-facing values, and hence are also taken to be pessimistic.

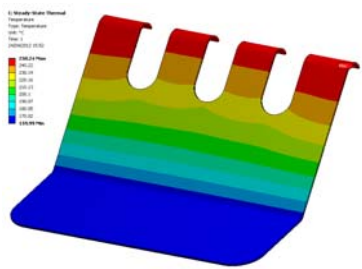

Figure 9: ANSYS thermal analysis results ($T_{max}$ = 250C)

The complex non-linear nature of the interaction between contact and crossbar meant that in order to verify that the mechanical behaviour of the system was well represented by the analysis and was practically viable a prototype test-rig was manufactured, as shown in figure 9.

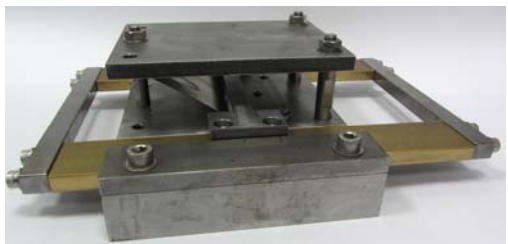

Fig 10: Mechanical prototype test rig

The assembly consisted of a plate onto which a 100mm wide spring, appropriately heat treated, was bolted, a sliding rail mechanism carrying a representative crossbeam, and a second plate, representing the vessel wall, mounted on a series of posts. The whole assembly was cycled to test for plastic deformation in the spring, actuation force needed, and contact force between the spring and vessel wall. Slight work hardening in the Inconel led to performance in these areas exceeding that predicted by analysis and confidence in the robustness of the design was confirmed.

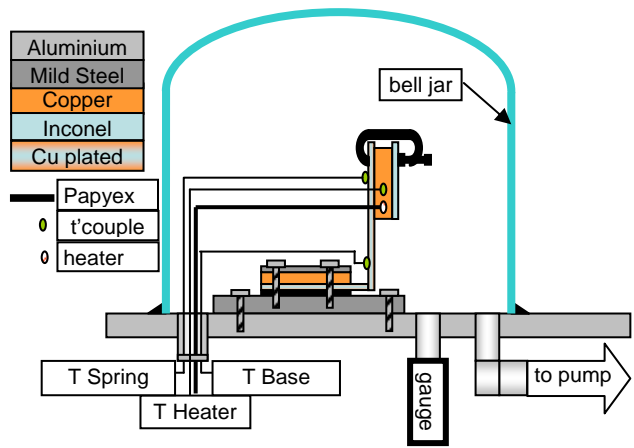

Fig 11: Thermal test rig

Due to the unpredictability of thermal conductance of bolted joints in vacuum, a test rig was developed to evaluate the effectiveness of the Papyex interlayer. This rig is shown in Figure 10. A temperature controlled heater element was attached to a plated Inconel strip, which was then bolted to a large steel base plate which acted as a heat sink. Thermocouples measured the temperature differences between heater and contact and between contact and steel plate. The whole assembly was placed under a bell jar and the ambient pressure reduced to ~$10^{-6}$ mbar. The heater temperature was raised to 250°C, the expected peak temperature from analysis. Without the Papyex interlayer, steady-state temperature differences >50°C were measured either side of the contact faces. Once included, the interlayer reduced these temperatures below the resolution of the equipment; though a slight time-dependent lag of order 5°C was noted. Temperatures equalised, however, within roughly ten minutes. Temperature gradients along the length of the spring were as predicted and input power required to reach 250°C matched analysis, showing an excellent thermal conductivity between contact and base plate.

Analysis had shown that the thickness of plating required was quite large (~100μm), and so it was necessary to ensure that the quality of coating achievable was sufficient. Preliminary trials have shown hardness and adhesion to be good up to thicknesses ~150μm, though surface quality for RF conductivity has been poor, primarily due to poor surface preparation and current control. Work is ongoing in this area.

## 5. Conclusions and next steps

A robust and simple RF grounding solution for the ITER ICRH has been designed, analysed, and tested. Optimisation of plating technique used and contact geometry will continue into the detailed design phase, aiming to ensure good RF performance and reliable electrical contact with the vessel wall. Since the preliminary design review in May 2012, values for disruption currents have been re-evaluated which are significantly larger than anticipated, causing large electromagnetic forces on the grounding contacts. While the concept outlined above is anticipated to be sound, a series of minor modifications are being explored which will enable the grounding system to survive these significantly increased loads.

## Acknowledgements

This work was funded partly by the RCUK Energy Programme under grant EP/I501045 and by F4E under grant 2009GRT-026. The views and opinions expressed herein do not necessary reflect those of F4E or European Commission or ITER Organisation. F4E is not liable for the use which might be made of the information in this publication.